\documentstyle[12pt]{article}
\def\eslt{E\llap/_T}
\def\to{\rightarrow}

\def\alt{\stackrel{<}{\sim}}

\def\tg{\tilde g}

\def\tnu{\tilde\nu}
\def\tell{\tilde\ell}
\def\tq{\tilde q}
\def\tt{\tilde t}
\def\tw{\widetilde W}
\def\tz{\widetilde Z}

\begin{document}
\begin{flushright}
UH-511-794-94
\end{flushright}
\begin{center}
{\large\bf SEARCHING FOR SUPERSYMMETRY:\\
A MINIREVIEW
\footnote{Presented at the Eighth Meeting of the Division of Particles and
Fields, Albuquerque, NM, August, 1994}}
\end{center}
\begin{center}
XERXES TATA \\
{\em Department of Physics and Astronomy, University of Hawaii},\\
{\em Honolulu, HI 96822, USA}
\end{center}
\date{}
\setlength{\baselineskip}{2.6ex}

\begin{center}
\parbox{13.0cm}
{\begin{center} ABSTRACT \end{center}
{\small \hspace*{0.3cm} After a lightning review of current bounds on the
masses of supersymmetric particles, we describe strategies that may be helpful
for extracting  signals from the production of squarks, gluinos or top
squarks, and from associated  chargino-neutralino production at the Tevatron.
We then briefly review SUSY signals at hadron and $e^+e^-$
supercolliders.
We discuss how various SUSY signals may be correlated within the supergravity
framework and indicate the sense in which $e^+e^-$ and hadron colliders may be
complementary.}}
\end{center}

\section {Current Status of Supersymmetry}

Aesthetic issues aside, the
interest in low energy SUSY phenomenology
was originally driven by the
recognition that SUSY can stablize the gauge hierarchy provided sparticles
are lighter than $O$(1 TeV). More recently,
the realization that the measured values of
gauge couplings at LEP are in agreement with the minimal supersymmetric
SU(5) model but {\it incompatible} with minimal non-supersymmetric
GUTs has led several groups to reexamine the expectations for sparticle
masses and mixing patterns within the theoretically appealing supergravity
(SUGRA) framework, the phenomenology of which we will return to in the last
section. Experimental constraints from flavour changing neutral currents can
readily be accommodated and, assuming $R$-parity is unbroken,
SUSY models include a viable candidate
for dark matter.

There is no direct evidence for sparticles in high energy collisions. This
is not (yet) a cause for despair since,
if we recall the original motivation for low energy SUSY, we would
typically expect sparticle masses of 100-1000~GeV. Since SUSY theories
are of the decoupling type in that virtual effects of sparticles decouple
as $m_{SUSY}\rightarrow\infty$, the
non-observation\cite{NODEV}
of any significant
deviations from the Standard Model (SM) in precision experiments at
LEP is compatible with (though not an argument for) low energy SUSY. The most
straightforward limits on masses of sparticles with gauge couplings to the $Z$
come from a measurement of its total width; these limits on the mass of
the slepton, the squark and the chargino ($\tw_1$), which are
independent of how sparticles decay, are a few GeV below
the kinematic bound $M_Z/2$. Exclusive searches and the measurement of
the invisible and leptonic widths of the $Z$ yield (sparticle
decay-dependent) lower limits
even closer to $M_Z/2$. Since
neutralinos ($\tz_i)$
couple to the $Z$ only via their Higgsino components, the bounds on their
masses are sensitive to model parameters which determine the mixing patterns.
The search
for squarks and gluinos, because of their strong interactions, is best
performed
at hadron colliders. The non-observation of an excess of $\eslt$ events
at the Tevatron has enabled\cite{TEVGL} the D0 collaboration to infer lower
limits of 150-160~GeV on their masses, improving on the earlier limit of
$\sim 100$~GeV obtained by CDF; if $m_{\tq}=m_{\tg}$, the mass bound improves
to 218~GeV. These bounds, which assume ten flavours of degenerate squarks,
have some sensitivity to the model parameters
which determine the cascade decay patterns of the $\tq$ and $\tg$. Within the
minimal supersymmteric model (MSSM), which is the commonly used framework for
experimental analyses, the Tevatron and LEP bounds together imply\cite{ROSZ}
a lower
bound just above 20~GeV on the mass of the $\tz_1$ which, because $\tz_1$
is assumed to be the (stable)
lightest SUSY particle (LSP), may have some cosmological
significance.

\section{SUSY Search in the 1990's}

The LEP collider is expected to enter its second phase of operation
around the end of next year when its energy will be upgraded to 175-200~GeV.
Given a data sample of $O$(100 $pb^{-1}$), the
clean environment of $e^+e^-$ collisions will readily enable
experimentalists to search for charginos, sleptons, squarks and even
Higgs bosons with masses up to 85-90~GeV ($b$-tagging capability may
be necessary if $m_H \simeq M_Z$). The corresponding
mass reach for neutralinos
is sensitive to (model-dependent) mixing angles.

The D0 and CDF experiments are collectively expected to accumulate an
integrated luminosity in excess of 100 $pb^{-1}$ by the end of the current
Tevatron run. In addition to extending the $\eslt$ search region, the
large increase in the data sample should enable them to
({\it i}) perform gluino and squark searches
via multilepton events from their cascade
decays, ({\it ii}) search for $\tw_1\tz_2$ production via isolated
trilepton events free of jet activity, and ({\it iii}) search for the
lighter $t$-squark, $\tt_1$. Experimental analyses will be greatly facilited by
the recent incorporation\cite{ISAJET} of SUSY processes into ISAJET.
We note that the Tevatron is unlikely
to be able to detect sleptons beyond the range of LEP\cite{SLEP}.

{\it Multilepton Signals from Gluinos and Squarks.} The conventional $\eslt$
search for $\tg$ and $\tq$ is background limited. Even with an integrated
luminosity of 1 $fb^{-1}$ that should be available with the Main Injector
upgrade of the Tevatron, we anticipate a maximum reach of $\sim 270$~GeV
($\sim 350$~GeV) if $m_{\tq} >> m_{\tg}$ ($m_{\tq} \simeq m_{\tg}$) in this
channel\cite{KAMON}.
Heavy gluinos and squarks can also decay via the chargino and
$\tz_2$ modes which, unless suppressed by phase space,
frequently dominate the decays of $\tq_L$ and $\tg$. The subsequent leptonic
decays of the $\tw_1$ and $\tz_2$ yield events with hard jets accompanied by
1-3 isolated, hard leptons and $\eslt$. While there are substantial
backgrounds to $1\ell$ and $\ell^+\ell^-$ event topologies, the {\it physics}
backgrounds in the $\ell^{\pm}\ell^{\pm}$ and $3\ell$ channels are essentially
negligible, so that these search channels are essentially rate limited. These
channels, which at the Main Injector have a reach\cite{BKTGL} of 230-300~GeV
depending
on $m_{\tq}$ and $m_{\tg}$, provide complementary ways of searching for
gluinos and squarks at the Tevatron, and because they are free of SM
backgrounds, may even prove superior if gluinos and squarks are very heavy.

{\it Search for Isolated Trilepton Events.} Associated $\tw_1\tz_2$ production
which occurs by $s$-channel $W^*$ and $t$-channel $\tq$ exchanges, followed by
the leptonic decays of $\tw_1$ and $\tz_2$ results in isolated trilepton plus
$\eslt$ events, with hadronic activity only from QCD radiation. SM
backgrounds to the $3\ell + n_{jet}\leq 1$ signal are negligible,
assuming that $WZ$ events can be vetoed with
high efficiency by requiring $m_{\ell\bar{\ell}}\not=M_Z$ within experimental
resolution; a conclusive observation of a handful of such events would,
therefore,
be a signal for new physics. The branching fraction for leptonic decays
of $\tz_2$, and sometimes also of $\tw_1$, and hence, this signal,
is enhanced if $m_{\tell} << m_{\tq}$ and the $\tz_2\tz_1Z$ is dynamically
suppressed; this situation frequently occurs in SUGRA type models, if
$m_{\tq} \simeq m_{\tg}$. Preliminary analyses by the CDF and D0
experiments\cite{WINO}
(for large values of the Higgsino mass parameter, $\mu$)
are already competitive with bounds from LEP.
The experiments will soon explore\cite{BKTWIN,KAMON}
parameter ranges not accessible at LEP
and, under favourable circumstances, may be competitive with LEP II. Within
the MSSM framework, this reach translates to $m_{\tg} = 250-400$~GeV depending
on $m_{\tq},\mu$ and $\tan\beta$, the ratio of the two Higgs vacuum
expectation values.

{\it Searching for Top Squarks at the Tevatron.}
The large Yukawa interactions of the top family which mix $\tt_L$ and $\tt_R$
serve to reduce the mass of $\tt_1$, the lighter of the
two mass eigenstates. In fact, it is theoretically possible that
$\tt_1$ is essentially massless with other squarks and gluinos all too heavy
to be produced at the Tevatron.
The Tevatron lower limits on $m_{\tq}$, are derived assuming ten
degenerate squark flavours, and so are not applicable to $\tt_1$. Currently,
the best limit, $m_{\tt_1}\alt M_Z/2$, comes from LEP experiments; this bound
can be evaded if the stop mixing angle and $m_{\tt_1}-m_{\tz_1}$ are both
fine-tuned, a possibility we do not entertain here.

The signals from top squark production depend on its decay patterns. For
$m_{\tt_1}< 125$~GeV, the range of interest at the Tevatron,
$\tt_1$ decays via the loop-mediated mode, $\tt_1\to c\tz_1$
when the
tree level decay $\tt_1\to b\tw_1$ is kinematically forbidden\cite{HK}.
Stop pair production is then signalled by $\eslt$ events from its direct
decays to the LSP.
With a data sample of 100 $pb^{-1}$,
Tevatron experiments should be able\cite{BST}
to probe stop masses up to 80-100~GeV,
significantly beyond the present bounds. On the other hand,
the tree level chargino
mode  dominates stop decays whenever it is kinematically allowed.
The subsequent leptonic
decay of one (or both) of the charginos lead to single lepton (dilepton)
+ $b$-jet(s)+$\eslt$ events, very similar to those expected from $t\bar{t}$
pair production. Top production is thus a formidable
background to the stop signal\cite{GUN}.
Also, for $m_t=175$~GeV, $m_{\tt_1} = 100$~GeV
and $m_{\tw_1}=70$~GeV, stop events would contribute about
33\% (20\%) of the
CDF top signal\cite{TOP} in the $1\ell$ (dilepton) channel.
Special cuts need to be devised to separate stop
from top events.
Since
stops accessible at the Tevatron are considerably lighter than
$m_t$, and
because the chargino, unlike $W$, decays via three body modes into
a {\it massive} LSP, stop events are generally softer than top events. It has
been shown\cite{BST} that by requiring $m_T$($\ell\eslt$)$<45$~GeV
($p_T(\ell^+)+p_T(\ell^-)+\eslt<100$~GeV), in additon to other canonical
cuts, stops with masses up to about 100~GeV should be detectable
in the $1\ell$ (dilepton) channel by the
end of the current Tevatron run;
for the single lepton channel sufficient
$b$-tagging capability is also required.

\section{Supersymmetry at Supercolliders}

Direct searches at the Main Injector and LEP II will probe
sparticle masses between 80-300~GeV; even assuming MSSM mass patterns, the
chargino search, by inference, will
probe gluino masses up to about 400~GeV. Since the SUSY mass scale
could easily be 1~TeV, it will, unless sparticles have already been
discovered, be up to
supercolliders such as the LHC at CERN or an
$e^+e^-$ Linear Collider (*LC) to explore the remainder of the parameter
space.

In the $\eslt$ channel, the LHC can search\cite{LHC} for gluinos and squarks
with
masses between 300~GeV to larger than 1~TeV. It is instructive to note that
several multilepton signals {\it must} simultaneously be present\cite{BTW}
if any
$\eslt$ signal is to be attributed to squark and gluino production, though
the various
relative rates could be sensitive to the entire sparticle spectrum.
The rate for like-sign dilepton plus $\eslt$ events is enormous
for $m_{\tg} \leq 300$~GeV; this ensures there is no window between the
Tevatron and the LHC where gluino of the MSSM may escape detection\cite{BTW}.
Gluinos
and squarks may also be a source of high $p_T$ $Z + \eslt$ events at the
LHC. The LHC can also search for ``hadron-free'' trilepton events from
$\tw_1\tz_2$ production. Backgrounds from top quark production can be very
effectively suppressed\cite{BCPT} by requiring that the two hardest
leptons have the
same sign of charge. The signal becomes unobservable when the two-body
decays $\tz_2\to (Z$ or $H_{\ell}) + \tz_1$ become accessible. The dilepton
mass distribution in $\ell^+\ell^-\ell'$ events can be used to reliably
measure $m_{\tz_2}-m_{\tz_1}$.
Selectrons and smuons with masses up to 250~GeV (300~GeV
if it is possible to veto central jets with an efficiency of 99\%) should
also be detectable\cite{SLEP}. Finally, we note that
with an integrated luminosity
of 100 $fb^{-1}$ the $\gamma\gamma$ decays of scalar stoponium has been
argued to allow for the detection of $\tt_1$ with a mass up to 250~GeV,
assuming $\tt_1\to b\tw_1$ (and, perhaps, also $\tt_1 \to bW\tz_1$)
is kinematically forbidden\cite{DN}.

Charged sparticles (and sneutrinos, if $m_{\tnu} > m_{\tw_1}$)
should be readily detectable at the *LC. Unfortunately, most detailed $e^+e^-$
studies to date do not incorporate the cascade decays
(which will be incorporated into ISAJET 7.11) of sparticles.
The real
power of these machines, however, lies in the ability to
do precision experiments which can then be used to probe\cite{MUR} unified
models of interactions discussed in the next section. *LC is
also the optimal facility to study the Higgs sector of SUSY\cite{HIGGS}.

\section{Supergravity Phenomenology}

Supergravity GUT models, via specific assumptions about the symmetries
of interactions responsible for SUSY breaking, provide an economical
framework for phenomenology by relating the many SUSY
breaking parameters of the MSSM. These relations hold
at some ultra-high unification scale
$M_X$ where these symmetries are manifest and the physics is simple.
Complex sparticle mass and mixing patterns (recently incorportated
into ISAJET 7.10),
{\it along with the correct breaking of electroweak symmetry} emerge
when these parameters are renormalized down to the weak scale as required for
phenomenology. The model is completely specified by just four SUSY
parameters which may be taken to be the common values of
SUSY-breaking gaugino and scalar
masses and trilinear scalar couplings, all specified at $M_X$, together
with the Higgs sector parameter $\tan\beta$.
In particular, $\mu$ and
the pseudoscalar Higgs boson mass are determined.
It is remarkable that even the simplest SUGRA GUTs are
consistent with experimental constraints as well as
cosmology\cite{SUGRA}.

Since the phenomenology is determined in terms of just
four SUSY parameters, various SUSY
cross sections become
correlated. Different experimental analyses
from $e^+e^-$ and hadron
colliders
can thus be
consistently combined into a single framework\cite{BCMPT}; since various
searches frequently probe different parts of the parameter space
they often complement one another. It should, however, be remembered
that this framework depends on assumptions about physics at the unification
scale. For experimental analyses we, therefore,
suggest using SUGRA models to obtain
default values of MSSM input parameters, and then, to test the sensitivity of
the predictions on the assumed SUGRA relations. Observation
of sparticles would not only be a spectacular new discovery, but a measurement
of their properties, particularly at linear colliders (where, it has
been argued\cite{MUR}, it is possible to do precision measurements of
sparticle parameters),
would test various
SUGRA assumptions, and so, serve as a telescope to the unification scale.

\vspace{4mm}
\noindent {\bf Acknowledgements:} Collaborations
with H.~Baer, M.~Bisset, C-H.~Chen,
M.~Drees, R.~Godbole, J.~Gunion, C.~Kao, R.~Munroe, M.~Nojiri, F.~Paige,
S.~Protopopescu, J.~Sender and J.~Woodside are gratefully acknowledged. This
research was supported by the U.S. Dept. of Energy grant DE-FG-03-94ER40833.
\vspace{6mm}

\bibliographystyle{unsrt}

\end{document}